\documentclass[preprint2]{aastex}
\usepackage{natbib}
\usepackage{booktabs}
\usepackage{threeparttable}

\newcommand{\fexxi}{Fe \scriptsize{XXI} \normalsize}
\newcommand{\ci}{C \scriptsize{I} \normalsize}
\newcommand{\oi}{O \scriptsize{I} \normalsize}
\newcommand{\siiv}{Si \scriptsize{IV} \normalsize}
\newcommand{\si}{S \scriptsize{I} \normalsize}
\newcommand{\feii}{Fe \scriptsize{II} \normalsize}
\newcommand{\siii}{Si \scriptsize{II} \normalsize}

\begin{document}

\title{Explosive chromospheric evaporation driven by nonthermal electrons around one footpoint of a solar flare loop}

\author{D. Li\altaffilmark{1,2,3}, Z. J. Ning\altaffilmark{1},
        Y. Huang\altaffilmark{1}, and Q. M. Zhang\altaffilmark{1,2}}
\affil{$^1$Key Laboratory of Dark Matter and Space Astronomy, Purple
Mountain Observatory, CAS, Nanjing 210008, China} \affil{$^2$CAS Key
Laboratory of Solar Activity, National Astronomical Observatories,
Beijing 100012, China} \affil{$^3$Key Laboratory of Modern Astronomy
and Astrophysics (Nanjing University), Ministry of Education,
Nanjing 210023, China} \altaffiltext{3}{Correspondence should be
sent to: lidong@pmo.ac.cn.}

\begin{abstract}
We explore the temporal relationship between microwave/HXR emission
and Doppler velocity during the impulsive phase of a solar flare on
2014 October 27 (SOL2014-10-27), which displays a pulse on the light
curves in microwave (34~GHz) and hard X-ray (HXR, 25$-$50~keV) bands
before the flare maximum. Imaging observation shows that this pulse
mainly comes from one footpoint of a solar flare loop. The slit of
{\it Interface Region Imaging Spectrograph} ({\it IRIS}) stays at
this footpoint during this solar flare. The Doppler velocities of
\fexxi~1354.09~{\AA} and \siiv~1402.77~{\AA} are extracted from the
Gaussian fitting method. We find that the hot line of
\fexxi~1354.09~{\AA} (log$T\sim$7.05) in corona exhibits blue shift,
while the cool line of \siiv~1402.77~{\AA} (log$T\sim$4.8) in
transition region exhibits red shift, indicating explosive
chromospheric evaporation. The evaporative upflows along the flare
loop are also observed in the AIA 131~{\AA} image. To our knowledge,
this is the first report of chromospheric evaporation evidence from
both spectral and imaging observations in the same flare. Both
microwave and HXR pulses are well correlated with the Doppler
velocities, suggesting that the chromospheric evaporation is driven
by nonthermal electrons around this footpoint of a solar flare loop.

\end{abstract}


\section{Introduction}
In a typical solar flare, the released energy is about 10$^{30}$~erg
within dozens of minutes. Such huge energy is thought to be
transferred from the magnetic energy via reconnection, which is
thought to heat the plasma and to accelerate the bi-directional
nonthermal electrons through the solar chromosphere, transition
region and corona. This is well known as the standard solar flare
model. These accelerated electrons are guided by the newly
reconnected magnetic field lines. Some of the accelerated electrons
travel up to the inter-planetary space, and the others propagate
downward to the lower corona and upper chromosphere, where they lose
their energies through Coulomb collisions with the denser plasma.
Subsequently, two footpoint sources are often produced along the
flare loop legs in the hard X-ray (HXR) and microwave bands
\citep{Brown71,Asai06}, and the double ribbons are brightened in
H$\alpha$, ultraviolet (UV), and extreme-ultraviolet (EUV)
wavelengths \citep{Czaykowska99,Del06,Milligan15}. Meanwhile, the
local chromospheric material is rapidly heated up to $\sim$10~MK
\citep{Antonucci82}, resulting in a higher pressure, which drives
the chromospheric material upward into the corona along the solar
flare loops. Thus the hot plasma fills the solar flare loops in a
process called `chromospheric evaporation'
\citep{Brown73,Fisher85a,Fisher85b,Liu06,Ning10,Brosius15,Tian15,Young15,Zhang16,Lee17}.
At the same time, the overpressure also drives the denser plasma
downward into the lower chromosphere in a process called
`chromospheric condensation'
\citep{Fisher85a,Fisher85b,Teriaca06,Li15}.

Chromospheric evaporation is observed in multiple wavelengths, i.e.,
HXR, EUV/UV, and microwave bands. Imaging observations show that the
HXR emission tends to rise with the double footpoint sources along
the loop legs, eventually merging them into a single source at the
loop top \citep[e.g.,][]{Liu06,Ning10,Ning11,Nitta12}. This process
is thought to be the HXR signature of chromospheric evaporation. On
the other hand, the heated materials rise upward to the corona and
disturb the local plasma. This process causes the microwave emission
suddenly cut-off on the dynamic spectra. Observationally, a
high-frequency cut-off drifts to low-frequency is the radio
signature of chromospheric evaporation
\citep{Aschwanden95,Karlicky98,Ning09}. Joint observations from
spectra and images show that the blue shift of hot lines in corona
tends to appear at the outside of flare ribbon
\citep{Czaykowska99,Li04,Li15}.

Chromospheric evaporation is divided into two types, i.e.,
`explosive' and `gentle'. Explosive evaporation takes place if the
input energy flux exceeds a critical value of
$\sim$10$^{10}$~erg~cm$^{-2}$~s$^{-1}$ \citep{Fisher85a,Fisher85b}.
In spectroscopic observations, explosive evaporation is usually
accompanied by blue shift with high speed
($\sim$100$-$400~km~s$^{-1}$) in the hot lines from corona, and by
red shift with low speed ($\sim$10$-$40~km~s$^{-1}$) in the cool
lines from upper chromosphere and transition region
\citep{Feldman80,Ding96,Del06,Teriaca06,Milligan09,Veronig10,Chen10,Brosius13,Doschek13,Tian14,Brosius15,Brosius16,Zhang16}.
Observationally, the speed of blue shift is an order of magnitude
larger than that of red shift, which is due to the fact that the
density of the overlying corona is less than that of the underlying
lower chromosphere \citep{Fisher85a,Fisher85b,Milligan09,Doschek13}.
Gentle evaporation occurs when the input energy flux is smaller than
the critical value, and all the emission lines display blue shift
from chromosphere through transition region to corona
\citep{Milligan06,Brosius09,Li11}.

In the literatures, there are three mechanisms to drive
chromospheric evaporation. The first one is electron-driven, which
emphases that the nonthermal electrons accelerated by magnetic
reconnection play an important role in driving chromospheric
evaporation
\citep{Fisher85b,Brosius03,Milligan09,Tian14,Tian15,Li15,Zhang16}.
The second one focusses on the thermal energy
\citep{Fisher85a,Falewicz09}. While the third one is driven by the
energy deposition through Alfv\'{e}n waves \citep{Reep16}. The
correlation between \fexxi13540.9~{\AA} Doppler velocity and HXR
emission from the whole flare region has already been documented in
the literatures \citep{Li15,Tian15,Lee17}. In this paper, using the
observations from {\it Interface Region Imaging Spectrograph}
\citep[{\it IRIS},][]{Dep14}, {\it NoRH} \citep{Hanaoka94}, {\it
Reuven Ramaty High Energy Solar Spectroscopic Imager} \citep[{\it
RHESSI},][]{Lin02}, Atmospheric Imaging Assembly
\citep[AIA,][]{Lemen12} and Helioseismic and Magnetic Imager
\citep[HMI,][]{Schou12} aboard {\it Solar Dynamics Observatory}
({\it SDO}), we study the temporal relationship between the
microwave/HXR emission and Doppler velocity of the SOL2014-10-27
event to prove the nonthermal electrons driving chromospheric
evaporation around one footpoint of a solar flare loop.

\section{Data Analysis and Results}
\begin{figure}
\epsscale{1.0} \plotone{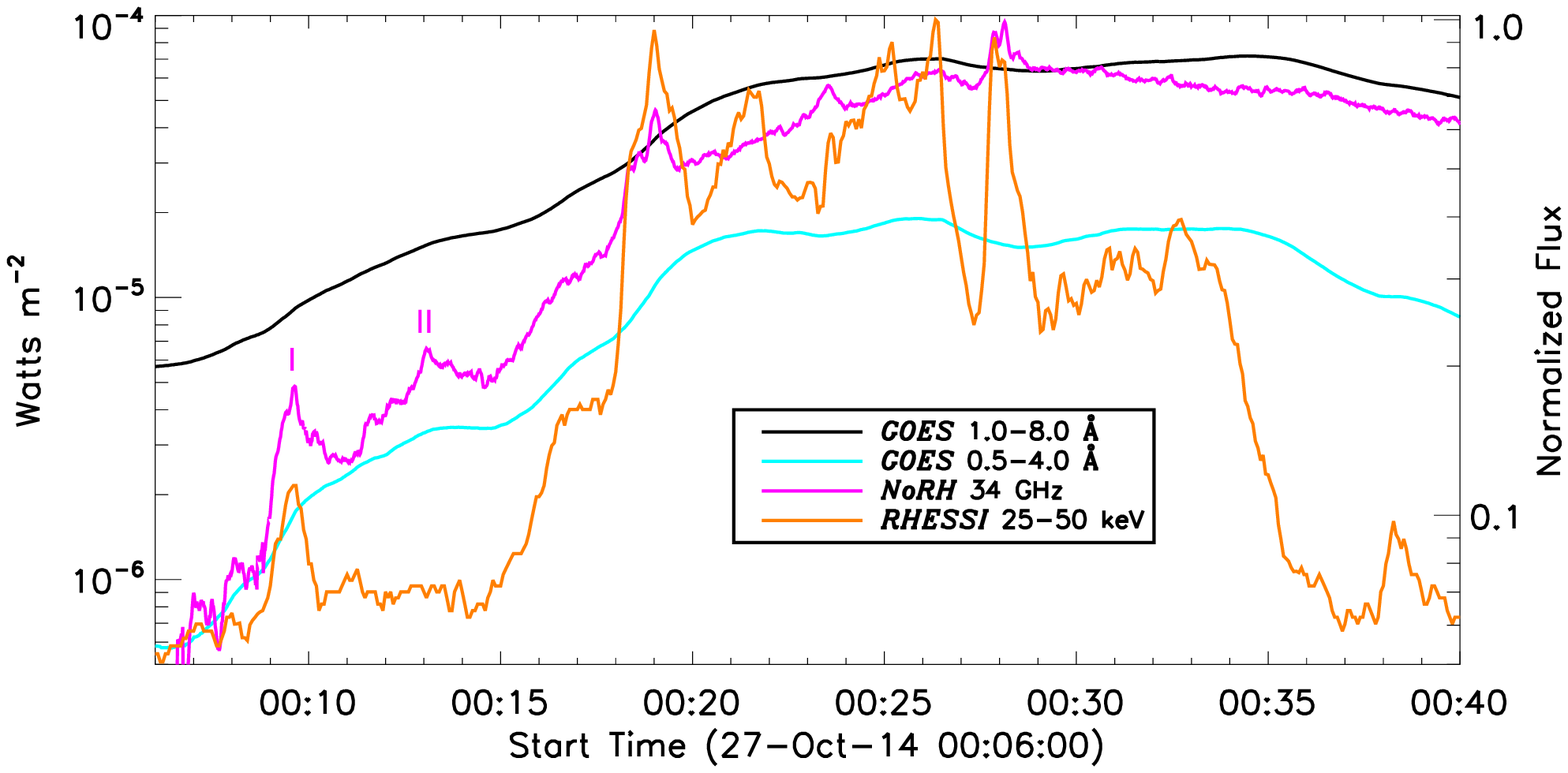} \caption{SXR light curves of 2014
October 27 flare from 00:06~UT to 00:40~UT in 1.0$-$8.0~{\AA}
(black) and 0.5$-$4.0~{\AA} (turquoise). Normalized flux in
microwave of 34~GHz (purple) and HXR 25$-$50~keV (orange) from the
whole flare region during the same time intervals. \label{flux}}
\end{figure}

The studied event takes place in NOAA AR12192 on 2014 October 27. It
is an M7.1 flare in {\it GOES} light curve. It displays typical
features of solar flares, such as two footpoint sources in microwave
and HXR images. The slit of {\it IRIS} stays at one footpoint during
the impulsive phase of the solar flare, which gives us a good chance
to study the temporal evolution of the Doppler velocity in coronal
and transition region lines, i.e., \fexxi~1354.09~{\AA} and
\siiv~1402.77~{\AA}. Figure~\ref{flux} shows the {\it GOES} SXR
fluxes in 1.0$-$8.0~{\AA} (black) and 0.5$-$4.0~{\AA} (turquoise),
respectively. The event starts at around 00:06~UT, and peaks at
about 00:34~UT from {\it GOES} SXR light curves. Meanwhile, the
microwave 34~GHz (purple) and HXR 25$-$50~keV (orange) emissions
display a series of regular and periodic peaks during the impulsive
phase of this solar flare, which result in several nonthermal pulses
\citep[e.g.,][]{Li17}, as marked by the Roman Capitals `I', `II'.
The first pulse (`I') is studied in this paper.

Figure~\ref{image} shows the multi-wavelength images from several
instruments during the first pulse. They have the same field-of-view
(FOV) of 90\arcsec$\times$90\arcsec. Panel~(a) displays the integral
intensity image ($\sim$2.45\arcsec/pixel) in microwave 34~GHz from
00:09~UT to 00:10~UT. The double footpoint sources are marked by the
white boxes. Among them, one is bright (ft1), while the other one is
faint (ft2). This is confirmed by {\it RHESSI} CLEAN image
($\sim$2\arcsec/pixel) at 25$-$50~keV, as shown in panel~(b). The
contour of ft2 is from the microwave emission. Panel~(c) gives the
line-of-sight (LOS) magnetogram ($\sim$0.6\arcsec/pixel) overlaid
with the AIA 1600~{\AA} contours. The bottom panels show the EUV/UV
snapshots in {\it SDO}/AIA 131~{\AA} and 1600~{\AA}
($\sim$0.6\arcsec/pixel), {\it IRIS}/SJI 1330~{\AA}
($\sim$0.33\arcsec/pixel). The AIA 1600~{\AA} image is applied to
co-align with the SJI 1330~{\AA} image \citep{Cheng15,Tian15,Li15},
because they both contain continuum emission around the temperature
minimum which is dominant in many bright features. The flare ribbons
in AIA 1600~{\AA} and SJI 1330~{\AA} are connected by the hot loops
in AIA 131~{\AA}, as indicated by the dashed lines. Consistent with
the standard solar flare model, panel~(e) shows that the double
footpoint sources overlap on the two flare ribbons. The slit of {\it
IRIS} crosses one flare ribbon and fixes on the edge of one
footpoint (ft1), as shown by the long blue line along the 45 degree
to the North-South direction.

\begin{figure}
\epsscale{1.0} \plotone{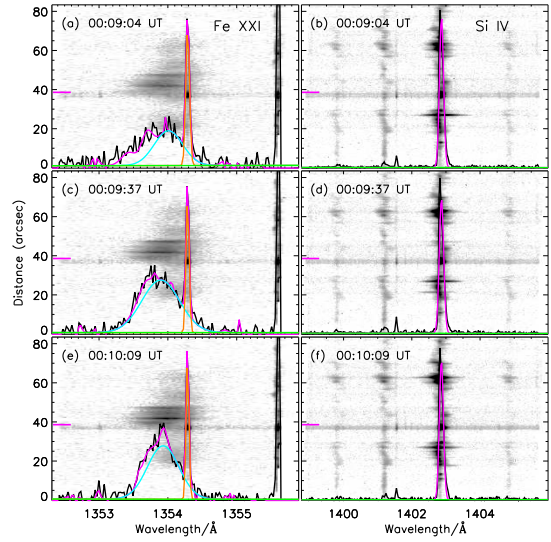} \caption{{\it IRIS} spectra at
`\fexxi' (left) and `\siiv' (right) windows on 2014 October 27. The
black profile is the line spectrum across the slit position marked
by a purple line on the left-hand side of each image, the
overplotted purple profile represents the fitting results. The
turquoise line is \fexxi~1354.09~{\AA}, and the orange line is
\ci~1354.29~{\AA}. The green line is the fitting background.
\label{spec}}
\end{figure}

The hot line of \fexxi~1354.09~{\AA}
\citep[log~$T$$\sim$7.05,][]{Dep14} in corona and the cool line of
\siiv~1402.77~{\AA} \citep[log~$T$$\sim$4.8,][]{Dep14} in transition
region are widely used to investigate chromospheric evaporation
\citep{Tian14,Tian15,Li15,Zhang16}. Figure~\ref{spec} shows the {\it
IRIS} observation at spectral windows of `\fexxi' (a, c, e) and
`\siiv' (b, d, f) at three times. The Y-axis is along the slit
direction. They have been pre-processed with the routines of
`iris\_orbitval\_corr\_l2.pro' \citep{Tian14,Cheng15} and
`iris\_prep\_despike.pro' \citep{Dep14} in Solar Soft Ware (SSW).
The absolute wavelength calibration is also manually performed with
the relatively strong neutral lines, i.e., `\oi'~1355.5977~{\AA} and
`\si'~1401.5136~{\AA} \citep[e.g.,][]{Tian15}. The overplotted black
profile shows the spectrum at the position of about 38.6\arcsec,
which is marked by the short purple line on the left-hand side. This
position is located at the edge of a flare ribbon in the northwest
labeled with the short blue line in Figure~\ref{image}~(e). The
multi-Gaussian functions (purple profile) superimposed on a linear
background (green line) are used to fit the {\it IRIS} spectrum at
`\fexxi' window to extract the flare line of \fexxi~1354.09~{\AA}.
The emission lines which are blended with the broad line of
\fexxi~1354.09~{\AA} should be excluded, i.e., \ci line at
1354.29~{\AA}, \feii lines at 1353.02~{\AA}, 1354.01~{\AA}, and
1354.75~{\AA}, \siii lines at 1352.64~{\AA} and 1353.72~{\AA}, some
unidentified lines at 1353.32~{\AA} and 1353.39~{\AA}
\citep{Polito15,Young15,Tian15}. Here, we fixed these blended line
positions, constrained their widths, and tied their intensities to
the lines in other spectral windows \citep[e.g.,][]{Li15,Li16}.
Finally, we can obtain the line profile of \fexxi 1354.09~{\AA}, as
shown by the turquoise line. The transition region line of
\siiv~1402.77~{\AA} is fitted with a single-Gaussian function
(purple profile) with a linear background (green line). Both the
Doppler velocities of \fexxi~1354.09~{\AA} and \siiv~1402.77~{\AA}
are detected by the fitting line centers subtracting the referenced
line centers \citep{Li14,Tian15}.

\begin{figure}
\epsscale{1.0} \plotone{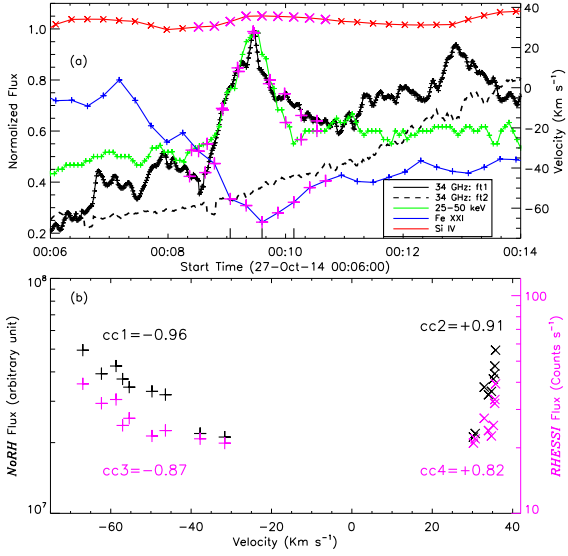} \caption{Panel~(a): The microwave
emission in {\it NoRP} 34~GHz at ft1 (solid black) and ft2 (dashed
black) during 00:06$-$00:14~UT. The HXR flux in {\it RHESSI}
25$-$50~keV (green) is from the whole flare at the same time. The
temporal evolution of Doppler velocities at \fexxi~1354.09~{\AA}
(blue) and \siiv~1402.77~{\AA} (red) lines are detected from a
distinct position along the slit of {\it IRIS}. Panel~(b): Scatter
plots of Doppler velocities dependence on microwave and HXR
emissions during the first pulse. The correlation coefficients (cc)
are labeled. \label{corr}}
\end{figure}

To exactly study the relationship between the microwave/HXR emission
and the evaporation speed, Figure~\ref{corr}~(a) shows the
normalized flux in microwave (black) and HXR (green) bands, and the
temporal evolution of Doppler velocity in \fexxi~1354.09~{\AA}
(blue) and \siiv~1402.77~{\AA} (red) on the slit position of
38.6\arcsec\ (short purple line in Figure~\ref{spec}). The hot line
of \fexxi~1354.09~{\AA} in corona displays blue shift, while the
cool line of \siiv~1402.77~{\AA} in transition region exhibits red
shift, which indicates the explosive chromospheric evaporation
around the footpoint of ft1. The maximum velocities of blue and red
shifts are about -67~km~s$^{-1}$ and 36~km~s$^{-1}$ during the first
pulse between 00:08$-$00:11~UT. The microwave light curves are solid
(black) at ft1 and dashed (black) at ft2. They are integrated over
the regions in the white boxes in Figure~\ref{image}~(a),
respectively. The observations show that the pulse (between
00:08$-$00:11~UT) studied here completely arise from ft1, and the
microwave emission at ft2 is monotonously increasing. However, the
HXR flux is integrated over the whole flare region, which is the
same as that in Figure~\ref{flux}. These light curves are well
correlated during the pulse interval, but their different time
cadences make the points impossible to correspond one-by-one. {\it
NoRH}, {\it RHESSI} and {\it IRIS} have the cadences of 1~s, 4~s and
16.2~s, respectively. In this paper, we use all 9 points from the
Doppler velocity around the pulse (between 00:08:30~UT and
00:11:00~UT), and their nearby points of microwave and HXR
emissions. Thus, 9 points are extracted from all the light curves,
as marked by the purple symbols (`+' or `$\times$').

Figure~\ref{corr}~(b) shows the Doppler velocities of
\fexxi~1354.09~{\AA} (`+') and \siiv~1402.77~{\AA} (`$\times$')
dependence on microwave 34~GHz (black) and HXR 25$-$50~keV (purple)
emissions during the pulse interval, i.e., between 00:08:30~UT and
00:11:00~UT. As expected from the model of electron-driven
chromospheric evaporation, we find a negative correlation
(-0.96/-0.87) between the microwave/HXR emission and the Doppler
velocity of hot line (\fexxi~1354.09~{\AA}) at ft1 (`-' indicates
that the Doppler velocity of \fexxi~1354.09~{\AA} is blue shift),
and a positive correlation (+0.91/+0.82) between the microwave/HXR
emission and the Doppler velocity of cool line (\siiv~1402.77~{\AA},
`+' indicates that the Doppler velocity of \siiv~1402.77~{\AA} is
red shift). Such high correlation coefficients suggest that the
nonthermal electrons which are accelerated by the magnetic
reconnection drive the explosive chromospheric evaporation around
the footpoint of ft1.

\begin{figure}
\epsscale{1.0} \plotone{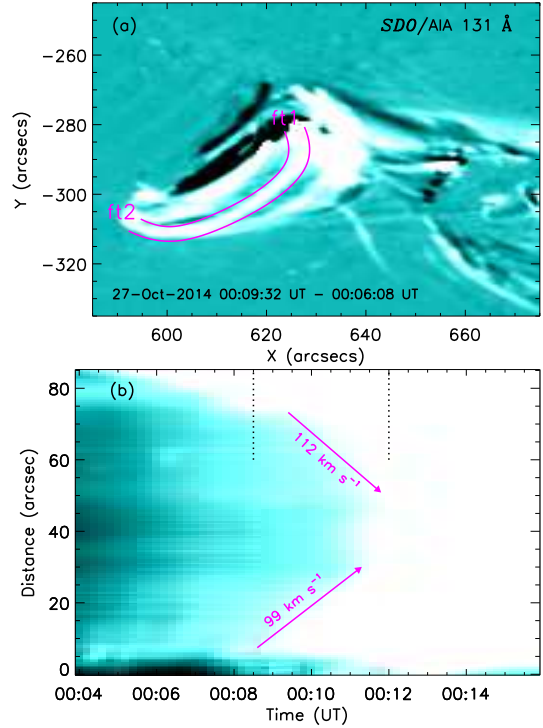} \caption{Upper: The difference map
in AIA 131~{\AA}, two solid lines (purple) outline a flare loop.
Bottom: The time-distance image along flare loop in AIA 131~{\AA}.
The start point (d=0) is from ft2. The arrows indicate the upflow
directions. Two dotted lines mark the upflow time between
00:08:30$-$00:12:00~UT. \label{slice}}
\end{figure}

Based on the chromospheric evaporation process, the plasma upflows
are expected to be observed along the flaring loops.
Figure~\ref{image}~(d) shows that the loops (red dashed lines) in
AIA 131~{\AA} connect the double ribbons in AIA 1600~{\AA}. These
loops become brighter and brighter with the flare development.
Meanwhile, these loops expand with time. Figure~\ref{slice}~(a)
shows the time difference image in AIA 131~{\AA} between the pulse
maximum ($\sim$00:09:32~UT) and the flare beginning
($\sim$00:06:08~UT). The two purple lines outline the expansion loop
which covers the hot plasma upflows due to chromospheric
evaporation. Figure~\ref{slice}~(b) gives its time-distance image,
and the Y-axis from bottom to top corresponds to the loop from ft2
to ft1. Thus the loop top is near the middle position in Y-axis,
i.e., $\sim$41\arcsec. It clearly shows the evaporation upflows from
double footpoints through legs to loop top, as indicated by the
purple arrows during 00:08:30$-$00:12:00~UT, which is consistent
with the microwave and HXR pulse (`I'). The evaporation speeds are
estimated to be $\sim$112~km~s$^{-1}$ and $\sim$99~km~s$^{-1}$
without considering the projection effect.

\section{Conclusions and Discussions}
Based on the observations from {\it IRIS}, {\it NoRH}, and {\it
RHESSI}, we explore the temporal relationship between Doppler
velocity of \fexxi~1354.09~{\AA} and \siiv~1402.77~{\AA} lines and
microwave/HXR emission in microwave 34~GHz and HXR 25$-$50~keV
during a pulse of solar flare on 2014 October 27. The completely new
observational result is that the explosive chromospheric evaporation
driven by nonthermal electrons originates from one footpoint of a
solar flare loop. Meanwhile, the evaporation upflows are also
observed from the double footpoints to loop top in AIA~131~{\AA}
image. This is consistent well with the standard solar flare model.
Our results agree well with previous findings about the temporal
correlation between the Doppler velocity and HXR flux which
represents the deposition rate \citep{Li15,Tian15,Lee17}, and are
also consistent well with the studies of the spatial correlation
between the upflows/dowflows and HXR sources in solar flares
\citep{Brosius092,Milligan09,Veronig10,Brosius16,Zhang16}. It has
been recently demonstrated that the explosive evaporation can also
be driven by the dissipation through Alfv\'{e}n waves, and this
mechanism can produce the HXR bursts \citep{Reep16}. Therefore, only
the presence of an HXR burst is impossible to rule out the
contribution from the heating of Alfv\'{e}n waves. On the other
hand, a microwave pulse is observed around the peak time of {\it
IRIS} Doppler velocity at one footpoint of a solar flare loop,
implying that the nonthermal electrons are injected into the
chromosphere along the flare loop in the Sun. Thus, the observations
in this case support an electron-driven chromospheric evaporation.

\begin{figure}
\epsscale{1.0} \plotone{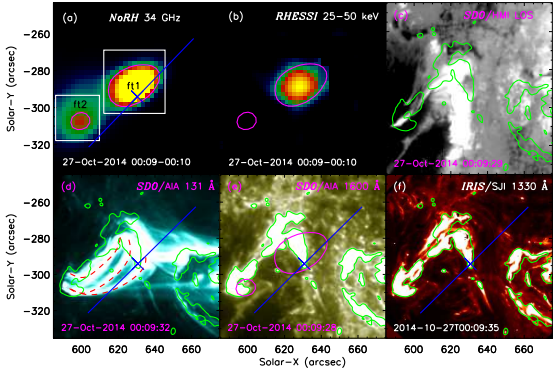} \caption{The multi-wavelength images
with a FOV of 90\arcsec$\times$90\arcsec on the 2014 October 27
flare. The purple contours are from the microwave of 34~GHz, and the
white boxes indicate the sources of two footpoints. The green
contours represent the AIA 1600~{\AA} intensities at the scale of
200~DN~s$^{-1}$. The red dashed lines in AIA~131~{\AA} outline the
possible flare loops. The long blue line represents the slit of {\it
IRIS} and the short blue line marks the region studied here.
\label{image}}
\end{figure}

Figure~\ref{image}~(e) and (f) shows that the slit of {\it IRIS}
crosses one of flare ribbons. The studied position of Doppler
velocity is just at the edge of flare ribbon, indicating that the
explosive chromospheric evaporation appears at the outside of flare
ribbon, which is consistent well with previous findings
\citep{Czaykowska99,Li04,Li15,Tian15}. On the other hand,
Figure~\ref{image}~(a) exhibits that the slit of {\it IRIS} does not
appear in the footpoint center, which may be due to the lower
spatial resolution of microwave image ($\sim$4.9$\arcsec$). Future
work needs higher resolution observations in microwave or HXR bands,
including the time and spatial resolutions.

In this letter, only the first pulse (`I') in microwave and HXR
emissions is used to investigate the explosive chromospheric
evaporation. We did not find any correspondence between the
microwave (or HXR) emission and Doppler velocity during the other
pulses, i.e., the second microwave pulse at about 00:13~UT (`II').
This may be because the upflows/downfows of the other microwave
pulses are very complex. The upflows are from the new heated plasma
which moves upward along the loop, while the downflows from the last
evaporated material may eventually cool and precipitate back down
along the loop \citep[e.g.,][]{Brosius03,Milligan09}. It could also
be because the location of energy deposition is not covered by the
slit of {\it IRIS}. The good correlation between Doppler velocity
and microwave/HXR emission may be found when shifting the slit
position \citep[e.g.,][]{Li15}.

\acknowledgments The authors would appreciate the anonymous referee
for his/her valuable comments and suggestions to improve the
manuscript. We thank the teams of {\it IRIS}, {\it GOES}, {\it
NoRH}, {\it RHESSI}, {\it SDO}/AIA and {it SDO}/HMI for their open
data use policy. The authors also thank Y. Gong for reading our
manuscript. This study is supported by NSFC under grants 11603077,
11573072, 11473071, 11333009, KLSA201708, and Laboratory No.
2010DP173032. Dr. D.~Li is also supported by the Youth Fund of
Jiangsu No. BK20161095, and Dr. Q.~M.~Zhang is supported by the
Surface Project of Jiangsu No. BK 20161618 and the Youth Innovation
Promotion Association CAS.

\end{document}